\documentclass[11pt,italian,a4widepaper]{article}

\usepackage{fancyhdr}            
\fancypagestyle{plain}{%
  \fancyhead{}                                     
  \fancyfoot[CE,CO]{\thepage}       
  \fancyhead[CE,CO]{\textcolor[rgb]{0.64,0.15,0.15}{Published in \emph{International Journal of Solids and Structures} \textbf{85-86}
  (2016), 67-75
\\
doi: http://dx.doi.org/10.1016/j.ijsolstr.2016.01.027}}
\setlength{\headheight}{14.5pt}}

\usepackage{latexsym}
\usepackage{amsmath,amsfonts,amssymb}
\usepackage{bm}
\usepackage{graphicx}
\usepackage{tabularx}
\usepackage{enumerate}
\usepackage{multirow}
\usepackage{booktabs}
\usepackage[font=footnotesize,labelfont=bf]{caption}
\usepackage{subfig}
\usepackage[svgnames,table]{xcolor}
\usepackage{latexsym}
\usepackage{amsmath}
\usepackage{booktabs}
\usepackage{multirow}
\usepackage{setspace}
\usepackage[toc,page]{appendix}
\usepackage{mathtools}
\usepackage{cancel}
\usepackage{relsize}
\usepackage{empheq}
\usepackage{mathrsfs}
\usepackage{breqn}
\usepackage{arcs}
\usepackage{amssymb}
\usepackage{wasysym}
\usepackage{pifont}

\begin{document}

\newcolumntype{L}[1]{>{\raggedright\let\newline\\\arraybackslash\hspace{0pt}}m{#1}}
\newcolumntype{C}[1]{>{\centering\let\newline\\\arraybackslash\hspace{0pt}}m{#1}}
\newcolumntype{R}[1]{>{\raggedleft\let\newline\\\arraybackslash\hspace{0pt}}m{#1}}

\def\ds{\displaystyle}

\newcommand{\beq}{\begin{equation}}
\newcommand{\eeq}{\end{equation}}
\newcommand{\lb}{\label}
\newcommand{\beqar}{\begin{eqnarray}}
\newcommand{\eeqar}{\end{eqnarray}}
\newcommand{\barr}{\begin{array}}
\newcommand{\earr}{\end{array}}
\newcommand{\jump}{\parallel}

\def\c{{\circ}}

\newcommand{\Ehat}{\hat{E}}
\newcommand{\That}{\hat{\bf T}}
\newcommand{\Ahat}{\hat{A}}
\newcommand{\chat}{\hat{c}}
\newcommand{\shat}{\hat{s}}
\newcommand{\khat}{\hat{k}}
\newcommand{\muhat}{\hat{\mu}}
\newcommand{\mc}{M^{\scriptscriptstyle C}}
\newcommand{\mei}{M^{\scriptscriptstyle M,EI}}
\newcommand{\mec}{M^{\scriptscriptstyle M,EC}}
\newcommand{\hbeta}{{\hat{\beta}}}
\newcommand{\rec}[2]{\left( #1 #2 \ds{\frac{1}{#1}}\right)}
\newcommand{\rep}[2]{\left( {#1}^2 #2 \ds{\frac{1}{{#1}^2}}\right)}
\newcommand{\derp}[2]{\ds{\frac {\partial #1}{\partial #2}}}
\newcommand{\derpn}[3]{\ds{\frac {\partial^{#3}#1}{\partial #2^{#3}}}}
\newcommand{\dert}[2]{\ds{\frac {d #1}{d #2}}}
\newcommand{\dertn}[3]{\ds{\frac {d^{#3} #1}{d #2^{#3}}}}

\def\bob{{\, \underline{\overline{\otimes}} \,}}
\def\ob{{\, \underline{\otimes} \,}}
\def\scalp{\mbox{\boldmath$\, \cdot \, $}}
\def\gdp{\makebox{\raisebox{-.215ex}{$\Box$}\hspace{-.778em}$\times$}}
\def\daa{\makebox{\raisebox{-.050ex}{$-$}\hspace{-.550em}$: ~$}}
\def\mK{\mbox{${\mathcal{K}}$}}
\def\cK{\mbox{${\mathbb {K}}$}}

\DeclarePairedDelimiter{\abso}{\lvert}{\rvert}
\DeclarePairedDelimiter{\norma}{\lVert}{\rVert}

\def\Xint#1{\mathchoice
   {\XXint\displaystyle\textstyle{#1}}%
   {\XXint\textstyle\scriptstyle{#1}}%
   {\XXint\scriptstyle\scriptscriptstyle{#1}}%
   {\XXint\scriptscriptstyle\scriptscriptstyle{#1}}%
   \!\int}
\def\XXint#1#2#3{{\setbox0=\hbox{$#1{#2#3}{\int}$}
     \vcenter{\hbox{$#2#3$}}\kern-.5\wd0}}
\def\ddashint{\Xint=}
\def\fpint{\Xint=}
\def\dashint{\Xint-}
\def\cpvint{\Xint-}
\def\intl{\int\limits}
\def\cpvintl{\cpvint\limits}
\def\fpintl{\fpint\limits}
\def\ointl{\oint\limits}
\def\bA{{\bf A}}
\def\ba{{\bf a}}
\def\bB{{\bf B}}
\def\bb{{\bf b}}
\def\bc{{\bf c}}
\def\bC{{\bf C}}
\def\bD{{\bf D}}
\def\bE{{\bf E}}
\def\be{{\bf e}}
\def\bbf{{\bf f}}
\def\bF{{\bf F}}
\def\bG{{\bf G}}
\def\bg{{\bf g}}
\def\bi{{\bf i}}
\def\bH{{\bf H}}
\def\bK{{\bf K}}
\def\bL{{\bf L}}
\def\bM{{\bf M}}
\def\bN{{\bf N}}
\def\bn{{\bf n}}
\def\b0{{\bf 0}}
\def\bo{{\bf o}}
\def\bX{{\bf X}}
\def\bx{{\bf x}}
\def\bP{{\bf P}}
\def\bp{{\bf p}}
\def\bQ{{\bf Q}}
\def\bq{{\bf q}}
\def\bR{{\bf R}}
\def\bS{{\bf S}}
\def\bs{{\bf s}}
\def\bT{{\bf T}}
\def\bt{{\bf t}}
\def\bU{{\bf U}}
\def\bu{{\bf u}}
\def\bv{{\bf v}}
\def\bw{{\bf w}}
\def\bW{{\bf W}}
\def\by{{\bf y}}
\def\bz{{\bf z}}
\def\T{{\bf T}}
\def\Te{\textrm{T}}
\def\Id{{\bf I}}
\def\bxi{\mbox{\boldmath${\xi}$}}
\def\balpha{\mbox{\boldmath${\alpha}$}}
\def\bbeta{\mbox{\boldmath${\beta}$}}
\def\bepsilon{\mbox{\boldmath${\epsilon}$}}
\def\bvarepsilon{\mbox{\boldmath${\varepsilon}$}}
\def\bomega{\mbox{\boldmath${\omega}$}}
\def\bphi{\mbox{\boldmath${\phi}$}}
\def\bsigma{\mbox{\boldmath${\sigma}$}}
\def\bfeta{\mbox{\boldmath${\eta}$}}
\def\bDelta{\mbox{\boldmath${\Delta}$}}
\def\btau{\mbox{\boldmath $\tau$}}
\def\tr{{\rm tr}}
\def\dev{{\rm dev}}
\def\div{{\rm div}}
\def\Div{{\rm Div}}
\def\Grad{{\rm Grad}}
\def\grad{{\rm grad}}
\def\Lin{{\rm Lin}}
\def\Sym{{\rm Sym}}
\def\Skw{{\rm Skew}}
\def\abs{{\rm abs}}
\def\Re{{\rm Re}}
\def\Im{{\rm Im}}
\def\capB{\mbox{\boldmath${\mathsf B}$}}
\def\capC{\mbox{\boldmath${\mathsf C}$}}
\def\capD{\mbox{\boldmath${\mathsf D}$}}
\def\capE{\mbox{\boldmath${\mathsf E}$}}
\def\capG{\mbox{\boldmath${\mathsf G}$}}
\def\tcapG{\tilde{\capG}}
\def\capH{\mbox{\boldmath${\mathsf H}$}}
\def\capK{\mbox{\boldmath${\mathsf K}$}}
\def\capL{\mbox{\boldmath${\mathsf L}$}}
\def\capM{\mbox{\boldmath${\mathsf M}$}}
\def\capR{\mbox{\boldmath${\mathsf R}$}}
\def\capW{\mbox{\boldmath${\mathsf W}$}}

\def\i{\mbox{${\mathrm i}$}}
\def\mC{\mbox{\boldmath${\mathcal C}$}}
\def\mB{\mbox{${\mathcal B}$}}
\def\mE{\mbox{${\mathcal{E}}$}}
\def\mL{\mbox{${\mathcal{L}}$}}
\def\mK{\mbox{${\mathcal{K}}$}}
\def\mV{\mbox{${\mathcal{V}}$}}
\def\C{\mbox{\boldmath${\mathcal C}$}}
\def\E{\mbox{\boldmath${\mathcal E}$}}

\def\ACME{{ Arch. Comput. Meth. Engng.\ }}
\def\ARMA{{ Arch. Rat. Mech. Analysis\ }}
\def\AMR{{ Appl. Mech. Rev.\ }}
\def\ASCEEM{{ ASCE J. Eng. Mech.\ }}
\def\acta{{ Acta Mater. \ }}
\def\CMAME {{ Comput. Meth. Appl. Mech. Engrg.\ }}
\def\CRAS{{ C. R. Acad. Sci., Paris\ }}
\def\EFM{{ Eng. Fract. Mech.\ }}
\def\EJMA{{ Eur.~J.~Mechanics-A/Solids\ }}
\def\IJES{{ Int. J. Eng. Sci.\ }}
\def\IJF{{ Int. J. Fracture}}
\def\IJMS{{ Int. J. Mech. Sci.\ }}
\def\IJNAMG{{ Int. J. Numer. Anal. Meth. Geomech.\ }}
\def\IJP{{ Int. J. Plasticity\ }}
\def\IJSS{{ Int. J. Solids Structures\ }}
\def\IngA{{ Ing. Archiv\ }}
\def\JAM{{ J. Appl. Mech.\ }}
\def\JAP{{ J. Appl. Phys.\ }}
\def\JE{{ J. Elasticity\ }}
\def\JM{{ J. de M\'ecanique\ }}
\def\JMPS{{ J. Mech. Phys. Solids\ }}
\def\Macro{{ Macromolecules\ }}
\def\MOM{{ Mech. Materials\ }}
\def\MMS{{ Math. Mech. Solids\ }}
\def\MMT{{ Metall. Mater. Trans. A}}
\def\MPCPS{{ Math. Proc. Camb. Phil. Soc.\ }}
\def\MRC{{ Mech. Res. Comm.}}
\def\MSE{{ Mater. Sci. Eng.}}
\def\PMPS{{ Proc. Math. Phys. Soc.\ }}
\def\PRE{{ Phys. Rev. E\ }}
\def\PRL{{ Phys. Rev. Letters\ }}
\def\PRSL{{ Proc. R. Soc.\ }}
\def\rock{{ Rock Mech. and Rock Eng.\ }}
\def\QAM{{ Quart. Appl. Math.\ }}
\def\QJMAM{{ Quart. J. Mech. Appl. Math.\ }}
\def\SCRMAT{{ Scripta Mater.\ }}
\def\SM{{\it Scripta Metall. }}

\def\salto#1#2{
[\mbox{\hspace{-#1em}}[#2]\mbox{\hspace{-#1em}}]}

\title{Isotoxal star-shaped polygonal voids and rigid inclusions in nonuniform antiplane shear fields.\\
Part I: Formulation and full-field solution}\date{}

\author{F. Dal Corso, S. Shahzad and D. Bigoni \\
DICAM, University of Trento, via Mesiano 77, I-38123 Trento, Italy }

\maketitle

\begin{abstract}
\noindent
An infinite class of nonuniform antiplane shear fields is considered for a linear elastic isotropic space and (non-intersecting)
isotoxal star-shaped polygonal voids and rigid inclusions perturbing these fields
are solved. Through the use of the complex potential technique together with the generalized binomial and the multinomial theorems,
full-field closed-form solutions are obtained in the conformal plane.
The particular (and important) cases of star-shaped cracks and rigid-line inclusions (stiffeners) are also derived.
Except for special cases (addressed in Part II), the obtained solutions  show singularities at the inclusion corners and at the crack and stiffener ends,
where the stress blows-up to infinity,  and is therefore detrimental to strength. It is for this reason that
the closed-form determination of the stress field near a sharp inclusion or void is crucial for the design of ultra-resistant composites.

\end{abstract}


\section{Introduction}

The investigation of the perturbation induced by an inclusion (a void, or a crack, or a stiff insert) in an ambient stress field loading a
linear elastic infinite space is a fundamental problem in solid mechanics, whose importance need not be emphasized. Usually this problem is
analyzed with respect to uniform ambient stress fields \cite{andersson, barbieri, chen2, jock, movmov, muskio}, although inhomogeneous, self-equilibrated stresses have
also been considered \cite{bacca, big1, das, schiavone, sen, van, vasudevan}. The interplay between stress inhomogeneities and singularities
generated at inclusion corners is important in the design of ultra-resistant composites, as stress singularities are known to be detrimental
to strength. In fact, an extreme stress concentration, leading to material failure, has been shown by experiments to
represent the counterpart of the mathematical concept of singularity \cite{inclusioni, noselli}.
The determination of the conditions leading to stress relief around inclusions may introduce new perspectives in the development of composite materials.

The present article addresses the analytical, closed-form solution of isotoxal star-shaped polygonal voids and rigid inclusions
in an elastic isotropic matrix loaded by inhomogeneous (but self-equilibrated) antiplane shear fields
(which are introduced as polynomial in an explicit mechanical setting).
The solution is obtained using the complex potential technique, with conformal mapping \cite{movmov,movchan,muskio},
which leads to a full-field determination of the stress field.
The particular cases of infinitely thin star corners are also addressed, corresponding to star-shaped cracks and stiffeners
(the latter also referred to as rigid-line inclusions).
These patterns of multiple cracks are quite common, as three and four point star-shaped cracks are induced by triangular and
Vickers pyramidal indenters \cite{chen, andrew, nazil}
and can emerge during drying of colloidal suspensions in capillary tubes \cite{gau, maurini}.
Multiple radial crack patterns are generated after low speed impacts\footnote{High speed
generates circumferential fractures in addition to radial.}  on brittle plates \cite{Vandenberghe}.
In Section \ref{caaaazzzooo}, using the multinomial (and the generalized binomial) theorem, the full-field closed-form
solutions for isotoxal star-shaped polygonal voids
and rigid inclusions (and for star-shaped cracks and stiffeners) perturbing an inhomogeneous antiplane shear field are
obtained, after the problem is posed and solved in its asymptotic form in Section \ref{duepa}.
These results open the way to issues related to inclusion neutrality and in particular allows the discovery of \lq quasi-static invisibility'
and \lq stress annihilations', whose treatment is deferred to Part II of this study \cite{partII}, together with considerations of irregularities
in the shape of the inclusions and the finiteness of the domain containing the inclusion.

The presented results, obtained in out-of-plane elasticity, provide also a solution for problems in thermal conductivity and electrostatics,
due to the common governing equations expressed by the Laplacian.

\section{Governing equations, polynomial far-field stress, and asymptotics} \lb{duepa}

When anti-plane strain conditions prevail in a linear elastic solid, the gradient of the only non-vanishing displacement component, orthogonal to the $x_1$--$x_2$ plane and denoted by
$w(x_1, x_2)$, defines the shear stress components (through the shear modulus $\mu$) as
\beq\label{g2}
\tau_{13}(x_1,x_2)=\mu\frac{\partial w(x_1,x_2)}{\partial x_1}, \qquad
\tau_{23}(x_1,x_2)=\mu\frac{\partial w(x_1,x_2)}{\partial x_2} ,
\eeq
which are requested to satisfy the equilibrium equation in the absence of body forces,
\beq\label{g1}
\frac{\partial \tau_{13}(x_1,x_2)}{\partial x_1}+\frac{\partial \tau_{23}(x_1,x_2)}{\partial x_2}=0.
\eeq
In addition to the equilibrium equations, compatibility (or, in other words, the Schwarz theorem for function $w$) requires
\beq
\label{compat}
\frac{\partial \tau_{13}(x_1,x_2)}{\partial x_2}-\frac{\partial \tau_{23}(x_1,x_2)}{\partial x_1}=0.
\eeq

Note that for the antiplane problem, one eigenvalue of the stress tensor is null and the other two have opposite signs.
The absolute value of the two non-null eigenvalues is
\beq\label{tautautau}
\tau=\sqrt{\left(\tau_{13}\right)^2+\left(\tau_{23}\right)^2}.
\eeq

\subsection{An infinite class of antiplane shear fields}

\begin{figure}[tp]
  \begin{center}
\includegraphics[width=8 cm]{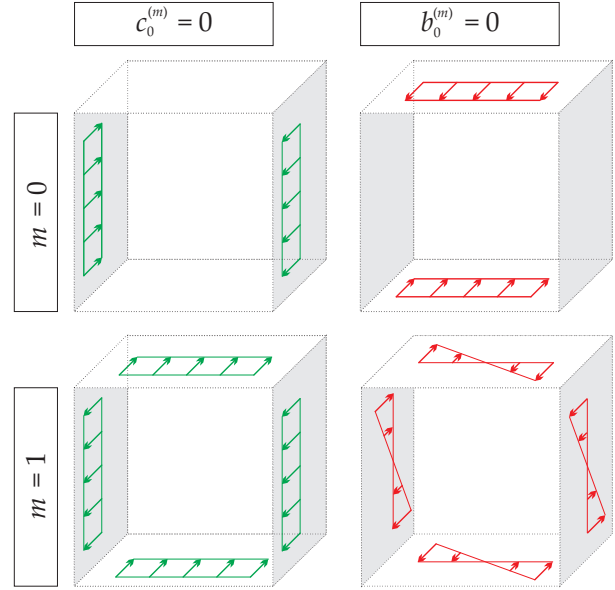}
\caption{The cases of uniform ($m=0$, upper part) and linear ($m=1$, lower part) remote (self-equilibrated) loading conditions
described by the constant $b_0^{(m)}$ (left) and $c_0^{(m)}$ (right), eqn (\ref{eq_general_b.c.}).
}
\label{fig_loading}
 \end{center}
\end{figure}

A class of remote anti-plane loadings is considered for an infinite elastic solid containing an inclusion, as defined by the following polynomial stress field of $m$-th order ($m \in \mathbb{N}$)
    \beq
        \label{eq_general_b.c.}
            \tau_{13}^{\infty (m)}(x_{1},x_{2})=
            \mathlarger{\sum_{j=0}^{m}} b_j^{(m)} x_1^{m-j}x_2^{j},\qquad
                        \tau_{23}^{\infty (m)}(x_{1},x_{2})=
            \mathlarger{\sum_{j=0}^{m}} c_j^{(m)} x_1^{m-j}x_2^{j},
                \eeq
where $b^{(m)}_{j}$ and $c^{(m)}_{j}$ are constants ($j=0,..., m$). Because the polynomial stress field (\ref{eq_general_b.c.}) has to satisfy both
the equilibrium equation (\ref{g1}) and the compatibility equation (\ref{compat}), all the constants
$b^{(m)}_{j}$ and $c^{(m)}_{j}$ are linearly dependent on $b^{(m)}_{0}$ and $c^{(m)}_{0}$  as follows

        \beq\begin{array}{llll}
        &b_{j}^{(m)}= (-1)^{\frac{j}{2}} \dfrac{m!}{j!(m-j)!}b_0^{(m)},\qquad
        &c_{j}^{(m)}= (-1)^{\frac{j}{2}} \dfrac{m!}{j!(m-j)!}c_0^{(m)}\,,\,\,
        &\forall\,\, \mbox{even}\,\, j\in[0;m],
        \\[5mm]
        &b_{j}^{(m)}= (-1)^{\frac{j-1}{2}} \dfrac{m!}{j!(m-j)!}c_0^{(m)},\qquad
        &c_{j}^{(m)}= (-1)^{\frac{j+1}{2}} \dfrac{m!}{j!(m-j)!}b_0^{(m)}\,,\,\,
        &\forall\,\, \mbox{odd}\,\, j\in[0;m].
        \end{array}
        \eeq
Note that the constants $b^{(m)}_{0}$ and $c^{(m)}_{0}$ represent a measure of the remote (or, in the inclusion problem,  the \lq unperturbed')
stress state along the $x_1$ axis,
\beq\label{guardaunpo}
\tau_{13}^{\infty (m)}(x_{1},0)= b_0^{(m)} x_1^{m},\qquad
\tau_{23}^{\infty (m)}(x_{1},0)= c_0^{(m)} x_1^{m},
\eeq
so that $b_0^{(0)}$ and $c_0^{(0)}$ are the loading constants defining the usual uniform Mode III, Fig. \ref{fig_loading} (upper part),
\beq
\tau_{13}^{\infty (0)}(x_{1},0)= b_0^{(0)},\qquad
\tau_{23}^{\infty (0)}(x_{1},0)= c_0^{(0)}.
\eeq
With the only exception of the case $m=0$, the two remote shear stress components are affected by both constants
$b_0^{(m)}$ and $c_0^{(m)}$. For example, in the case of linear remote loading ($m$=1), Fig. \ref{fig_loading} (lower part), the remote field
is defined by
\beq\label{guardaunpo}
\tau_{13}^{\infty (1)}(x_{1},x_2)= b_0^{(1)} x_1+c_0^{(1)} x_2,\qquad
\tau_{23}^{\infty (1)}(x_{1},x_2)= c_0^{(1)} x_1- b_0^{(1)} x_2.
\eeq
Note that the introduced polynomial fields can be used to reconstruct through series a general self-equilibrated remote loading.
Therefore, due to the superposition principle,
the solution obtained in the next sections also describes the mechanical fields under general Mode III remote loadings.

It is instrumental to express the polynomial stress field (\ref{eq_general_b.c.}) in two further reference systems, one Cartesian and the other polar.
In particular, with reference to a $\hat{x}_1$--$\hat{x}_2$ Cartesian coordinate system obtained through a rotation of an angle $\gamma$ of a $x_1$--$x_2$ system,
the polynomial stress field can be expressed as
    \beq
        \label{eq_general_b.c.2}
            \hat{\tau}_{13}^{\infty (m)}(\hat{x}_{1},\hat{x}_{2})=
            \mathlarger{\sum_{j=0}^{m}} \hat{b}_j^{(m)} \hat{x}_1^{m-j}\hat{x}_2^{j},\qquad
                        \hat{\tau}_{23}^{\infty (m)}(\hat{x}_{1},\hat{x}_{2})=
            \mathlarger{\sum_{j=0}^{m}} \hat{c}_j^{(m)} \hat{x}_1^{m-j}\hat{x}_2^{j},
                \eeq
where the loading constants $\hat{b}^{(m)}_0$ and $\hat{c}^{(m)}_0$ are linearly dependent on the constants
 $b^{(m)}_0$ and $c^{(m)}_0$ as follows
\beq
\label{eq_rotaz}
\begin{array}{ll}
\hat{b}^{(m)}_0 = b_0^{(m)} \cos((m+1)\gamma) + c_0^{(m)} \sin((m+1)\gamma),\\[4mm]
\hat{c}^{(m)}_0 = c_0^{(m)} \cos((m+1)\gamma) - b_0^{(m)} \sin((m+1)\gamma).
\end{array}
\eeq
With reference to a polar coordinate system $(r, \theta)$ centered at the origin of the $x_1$--$x_2$ axes,
the polynomial stress field (\ref{eq_general_b.c.}) can be rewritten as
    \beq
        \begin{cases}
        \barr{cll}
        \label{eq_general_b.c.polar}
            \tau_{r 3}^{\infty (m)}(r, \theta)&=&
            r^{m} \left[b^{(m)}_0 \cos\left((m+1)\theta\right)+c^{(m)}_0 \sin\left((m+1)\theta\right)\right],\\[6mm]
            \tau_{\theta 3 }^{\infty (m)}(r, \theta)&=&
            r^{m} \left[c^{(m)}_0 \cos\left((m+1)\theta\right)-b^{(m)}_0 \sin\left((m+1)\theta\right)\right],
                \earr
        \end{cases}
    \eeq
corresponding to the displacement
    \beq
        \label{eq_general_b.c.polar2}
            w^{\infty (m)}(r, \theta)=\frac{r^{m+1}}{\mu (m+1)}
             \left[b^{(m)}_0 \cos\left((m+1)\theta\right)+c^{(m)}_0 \sin\left((m+1)\theta\right)\right].
    \eeq
Finally, it can be noted that the modulus of the principal stress (\ref{tautautau}) is independent of the circumferential angle $\theta$
    \beq
\label{eigeninfinity}
            \tau^{\infty (m)}(r)=
            r^{m} \sqrt{\left[b^{(m)}_0\right]^2+\left[c^{(m)}_0\right]^2},
    \eeq
so that the level sets of the modulus  of the (unperturbed) shear stress are concentric circles centered at the origin of the axes.

\subsection{Asymptotic expansion near the vertex of a void or a rigid inclusion}

\begin{figure}[tp]
  \begin{center}
\includegraphics[width=12 cm]{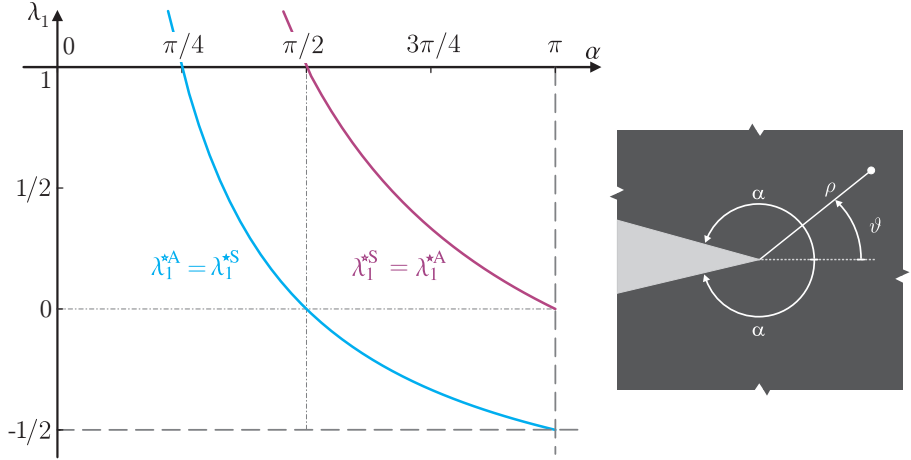}
\caption{
(left) First eigenvalue $\lambda_1$, guiding the leading-order term in the asymptotic description of
symmetric and antisymmetric parts of mechanical fields (\ref{espandi}) in the neighborhood of the inclusion vertex,
as a function of the semi-angle  $\alpha$ exterior to the inclusion (right).
} \label{fig_singularita}
 \end{center}
\end{figure}

A vertex of a void or a rigid inclusion is considered (defined by the semi-angle $\alpha$ exterior to the inclusion, Fig. \ref{fig_singularita}, right), with reference to a polar coordinate
system $(\rho,\vartheta)$ centered at the inclusion corner, where
$\vartheta\in(-\alpha,\alpha)$ measures the angle from the symmetry axis.
Following \cite{roaz1}--\cite{roaz5}, the solution of the general out-of-plane problem can be decomposed in
its symmetric $w^{\textup{S}}(\rho,\vartheta)=w^{\textup{S}}(\rho,-\vartheta)$ and antisymmetric
$w^{\textup{A}}(\rho,\vartheta)=-w^{\textup{A}}(\rho,-\vartheta)$ terms,
\beq
w(\rho,\vartheta)=w^{\textup{S}}(\rho,\vartheta)+w^{\textup{A}}(\rho,\vartheta),
\eeq
which, considering equations (\ref{g2})--(\ref{compat}),
assume the following expressions in polar coordinates,
\beq\begin{array}{lll}\label{espandi}
w^{\textup{S}}(\rho,\vartheta)= \ds D^{\textup{S}} \rho^{1+\lambda^{\textup{S}}} \cos\left[\left(1+\lambda^{\textup{S}}\right)\vartheta\right],\\[4mm]
w^{\textup{A}}(\rho,\vartheta)= D^{\textup{A}} \ds\rho^{1+\lambda^{\textup{A}}} \sin\left[\left(1+\lambda^{\textup{A}}\right)\vartheta\right],
\end{array}
\eeq
and, through the isotropic elastic constitutive relation (\ref{g2}), the following stress field representations are obtained as
\beq\begin{array}{lll}
\label{espandi2}
\tau_{\rho3}^{\textup{S}}(\rho,\vartheta)= \mu\ds \, D^{\textup{S}} \,(1+\lambda^{\textup{S}})
\rho^{\lambda^{\textup{S}}} \cos\left[(1+\lambda^{\textup{S}})\vartheta\right], \\[4mm]
\tau_{\vartheta3}^{\textup{S}}(\rho,\vartheta)= - \mu \ds \, D^{\textup{S}} \,(1+\lambda^{\textup{S}})
\rho^{\lambda^{\textup{S}}} \sin\left[(1+\lambda^{\textup{S}})\vartheta\right],\\[4mm]
\tau_{\rho3}^{\textup{A}}(\rho,\vartheta)= \mu\ds \,  D^{\textup{A}} \,(1+\lambda^{\textup{A}})
\rho^{\lambda^{\textup{A}}} \sin\left[(1+\lambda^{\textup{A}})\vartheta\right], \\[4mm]
\tau_{\vartheta3}^{\textup{A}}(\rho,\vartheta)=\mu\ds\, D^{\textup{A}} \,(1+\lambda^{\textup{A}})
\rho^{\lambda^{\textup{A}}} \cos\left[(1+\lambda^{\textup{A}})\vartheta\right],
\end{array}
\eeq
where $D^{\textup{S}}$ and $D^{\textup{A}}$ are constants (to be defined in relation to the remote loading), while $\lambda^{\textup{S}}$ and
$\lambda^{\textup{A}}$ are the eigenvalues of the characteristic equations for the symmetric and antisymmetric problem, respectively,
with $\{\lambda^{\textup{S}},\lambda^{\textup{A}}\}>-1$, to satisfy the requirement of finiteness of the local elastic strain energy.
These eigenvalues can be defined through the boundary condition imposed at the inclusion boundary and are crucial to the
asymptotic description of stress fields around the inclusion vertex.

The apexes $^{\text{\tiny\ding{73}}}$ and $^{\text{\tiny\ding{72}}}$ will be used to distinguish between values
assigned to voids and to rigid inclusions, respectively.
The null traction or null displacement boundary conditions at $\theta =\pm \alpha$, holding respectively for the former and the latter problem,
can be expressed as \cite{andrzei}
\beq
\tau^{\text{\tiny\ding{73}}}_{\theta 3}(\rho,\pm \alpha)=0,\qquad
w^{\text{\tiny\ding{72}}} (\rho,\pm \alpha)=0,
\eeq
leading to the following characteristic equations
\beq\begin{array}{lll}
 \sin\left[\alpha\left(1+\lambda^{\!\text{\tiny\ding{73}}\textup{S}}\right)\right]=0, \qquad
 \cos\left[\alpha\left(1+\lambda^{\!\text{\tiny\ding{73}}\textup{A}}\right)\right]=0,\\[6mm]
 \cos\left[\alpha\left(1+\lambda^{\!\text{\tiny\ding{72}}\textup{S}}\right)\right]=0,
 \qquad  \sin\left[\alpha\left(1+\lambda^{\!\text{\tiny\ding{72}}\textup{A}}\right)\right]=0,
\end{array}
\qquad j \in \mathbb{N}, \eeq
and from which two countably infinite set of eigenvalues $ \lambda^{\text{\tiny\ding{73}}\textup{A}}_j$, $ \lambda^{\text{\tiny\ding{72}}\textup{A}}_j$, $ \lambda^{\text{\tiny\ding{73}}\textup{S}}_j$ and $ \lambda^{\text{\tiny\ding{72}}\textup{S}}_j$ are obtained as
\beq
\begin{cases}
\lambda^{\!\text{\tiny\ding{73}}\textup{S}}_j(\alpha)=\lambda^{\!\text{\tiny\ding{72}}\textup{A}}_j(\alpha)=-1 +  \dfrac{j \pi}{\alpha},        \\[6mm]
\lambda^{\!\text{\tiny\ding{73}}\textup{A}}_j(\alpha)=\lambda^{\!\text{\tiny\ding{72}}\textup{S}}_j(\alpha)=-1 +
\dfrac{(2j-1) \pi}{2\alpha},
\end{cases}
\qquad j \in \mathbb{N}. \eeq
The mechanical fields at small distances from the inclusion are ruled by
the leading-order term in the symmetric and antisymmetric  asymptotic expansions (\ref{espandi}), which correspond to $j=1$
\beq
\begin{cases}
\lambda^{\!\text{\tiny\ding{73}}\textup{S}}_1(\alpha)=\lambda^{\!\text{\tiny\ding{72}}\textup{A}}_1(\alpha)=-1 +  \dfrac{\pi}{\alpha}\geq 0, \\[5mm]
\lambda^{\!\text{\tiny\ding{73}}\textup{A}}_1(\alpha)=\lambda^{\!\text{\tiny\ding{72}}\textup{S}}_1(\alpha)=-1
+  \dfrac{\pi}{2\alpha}\geq \ds-\frac{1}{2},
\end{cases}
\eeq
and are reported  in Fig. \ref{fig_singularita} (left) as a function of exterior semi-angle $\alpha$. Note that the following property holds true
\beq
\lambda^{\!\text{\tiny\ding{73}}\textup{S}}_1(\alpha)=\lambda^{\!\text{\tiny\ding{72}}\textup{A}}_1(\alpha)\,\,>
\,\,\lambda^{\!\text{\tiny\ding{73}}\textup{A}}_1(\alpha)=\lambda^{\!\text{\tiny\ding{72}}\textup{S}}_1(\alpha).
\eeq

The range of variation of the values $\lambda_1$ for different values of the
exterior semi-angle $\alpha$ is summarized in Tab. 1  and reported in Fig. \ref{fig_singularita} (left), from which it
can be noted that the stress field has:
\begin{itemize}
\item a singular leading-order term for  antisymmetric notch/symmetric wedge problems when
$ \alpha \in \left(\pi/2,\pi\right]$ (in particular a square-root
singularity is attained for $\alpha=\pi$, corresponding to antisymmetric crack/symmetric stiffener problems);

\item a constant (zeroth-order) term for  antisymmetric notch/symmetric wedge problems when
$ \alpha =\pi/2$ and for  symmetric notch/antisymmetric wedge
problems when $ \alpha =\pi$. Such a constant is usually called
T-stress in in-plane and S-stress in out-of-plane \cite{Gupta,Moon, radaj} crack problems;

\item a non-singular leading-order term for symmetric notch/antisymmetric wedge problems when $ \alpha < \pi$ and
for antisymmetric notch/symmetric wedge problems when $\alpha <\pi/2$.
\end{itemize}

The above-listed observations are crucial for the understanding of the
occurrence of stress singularity or of stress annihilation at the
vertices of polygonal void and rigid inclusions, as shown in Part II.
\begin{table}
\begin{center}
\begin{tabular}{c|cccc}
\toprule[.8pt]
$\alpha  $& $\in \left(0, \dfrac{\pi}{2}\right) $&$\dfrac{\pi}{2}$&$\in \left(\dfrac{\pi}{2}, \pi\right) $&$\pi$\\[4mm]
\hline
\\
$\lambda^{\!\text{\tiny\ding{73}}\textup{S}}_1(\alpha)=\lambda^{\!\text{\tiny\ding{72}}\textup{A}}_1(\alpha)$&  $>1$ & 1 & $\in (0, 1)$ & 0\\[6mm]
$\lambda^{\!\text{\tiny\ding{73}}\textup{A}}_1(\alpha)=\lambda^{\!\text{\tiny\ding{72}}\textup{S}}_1(\alpha)$&  $>0$ & 0 & $\in \left(-\dfrac{1}{2}, 0\right)$ & $-\dfrac{1}{2}$\\
\bottomrule
\end{tabular}
\bf \small \caption{\textnormal{Ranges of the first eigenvalue
$\lambda_1$, defining the leading-order term in the description of
symmetric and antisymmetric parts of the displacement field $w$, eqn (\ref{espandi})$_1$, for
different ranges of exterior semi-angle  $\alpha$ (Fig \ref{fig_singularita}, right). }}
\end{center}
\end{table}

\section{Full-field solution}\lb{caaaazzzooo}

The full-field solution for non-intersecting isotoxal star polygonal voids and rigid inclusions embedded in an isotropic elastic material subject
to the generalized anti-plane remote polynomial
stress field (\ref{eq_general_b.c.}) is obtained through the complex potential technique generalizing the solution by Kohno and Ishikawa \cite{japan}.
Considering the constitutive relation (\ref{g2}), equilibrium in the absence of body-forces (\ref{g1}) can
be expressed in terms of the displacement field $w$ as the Laplace equation
\beq
\nabla^2 w\left(x_1,x_2\right)=0,
\eeq
so that, introducing a complex potential $f(z)$, function of the complex variable $z=x_1+\i x_2$ (where $\i$ is the imaginary unit), such that
\beq\label{pot1}
w=\frac{1}{\mu}\Re[f(z)],
\eeq
and, considering the Cauchy-Riemann conditions for analytic functions, the stress-potential relationship can be written as
\beq\label{pot2}
\tau_{13}-\i \tau_{23}=f'(z),
\eeq
so that the out-of-plane resultant shear force $F_{\stackrel\frown {B C}}$ along an arc $\stackrel{\frown}{B C}$ is
\beq\label{pot3}
F_{\stackrel{\frown}{BC}}=\mathlarger{\int^{C}_{B}}\left(\tau_{13}\,\mathrm{d}x_{2}-\tau_{23}\,\mathrm{d}x_{1}\right)
=\Im \left[f(z_{C})-f(z_{B})\right].
\eeq
The complex potential $f(z)$ can be considered as the sum of the unperturbed potential $f^{\infty}(z)$,
which is the solution in the absence of the inclusion, and the perturbed one, $f^{p}(z)$,
 introduced to recover the boundary condition along the inclusion boundary,
\beq
f(z)=f^{\infty}(z)+f^{p}(z).
\eeq
Using the polynomial description (\ref{eq_general_b.c.}) for the self-equilibrated remote stress field $\tau_{13}^{\infty (m)}$ and
$\tau_{23}^{\infty (m)}$, the unperturbed potential is given as
\beq
\label{eq_unperturbed_potential}
f^{\infty}(z,m)= T^{(m)} z^{m+1} \, ,
\eeq
where
\beq
\label{eq_unperturbed_potential_1}
T^{(m)}= \frac{b^{(m)}_0-\i  \, c^{(m)}_0}{m+1},
\eeq
which in the particular case of uniform antiplane shear load, $m=0$, reduces to \cite{japan} (their equation (20)).

Considering now the presence of the inclusion, it is instrumental to define the  conformal mapping
$z=\omega(\zeta)$, which transforms the boundary of the inclusion in the physical plane into a circle of unit radius within the
conformal plane.
In the conformal plane, the complex potential
\beq
\lb{pot}
g(\zeta)=f(\omega(\zeta)),
\eeq
is introduced, so that equations (\ref{pot1}), (\ref{pot2})
and (\ref{pot3}) become
\beq
\label{eq_transformed_fullfield}
w=\frac{1}{\mu}\Re[g(\zeta)], \qquad \tau_{13}-\i \tau_{23}=\frac{g'(\zeta)}{\omega'(\zeta)},
\qquad
F_{\stackrel\frown {BC}}=\Im \left[g(\zeta_{B})-g(\zeta_{C})\right].
\eeq
The displacement and stress fields in the physical domain can be obtained once the inclusion shape is specified.

\begin{figure}[tp]
  \begin{center}
\includegraphics[width=13 cm]{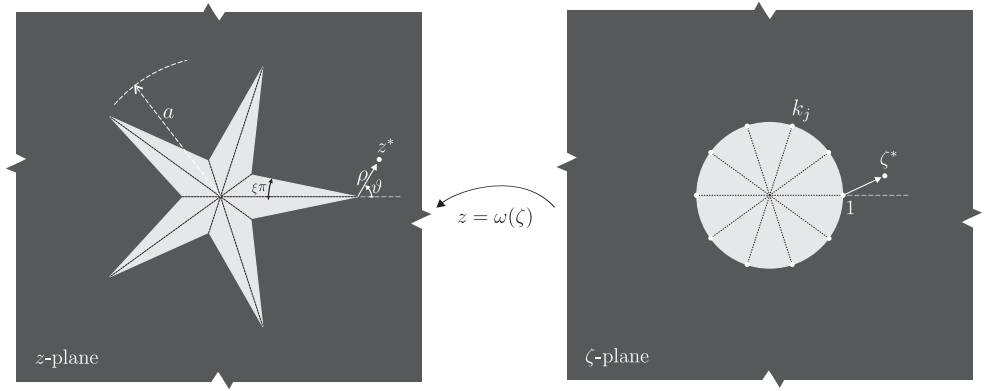}
\caption{Infinite plane containing a $n$-pointed isotoxal star-shaped polygon inscribed in a circle of radius $a$.
The polygon is defined by the semi-angle $\xi\pi$ at the isotoxal-points ($z$-plane) and is conformally
mapped onto an infinite plane with a circular inclusion of unit radius ($\zeta$-plane) using the
Schwarz-Christoffel formula (\ref{eq_sc_integral_ext_ext_genral}).
Note the local reference systems $z^{*} = z-a$ and $\zeta^{*} = \zeta - 1$ defined in the two planes.}
\label{fig_star_shaped_poly_mapping}
 \end{center}
\end{figure}

Rigid inclusions and voids are considered of isotoxal polygonal star shape, see Fig. \ref{fig_star_shaped_poly_mapping} (left), embedded in infinite elastic matrix. An isotoxal polygonal star is defined by the number $N$ of vertices
and a number $n=N/2$ of points. Note that $N \geq 4$ is always even, while $n \geq 2$ is an integer, so that a polygon (for instance a triangle, characterized by $N=6$ and $n=3$) is viewed as
a degenerate star (for instance a three-pointed star, having $N=6$ and $n=3$).
Introducing $\beta_j$ as the fraction of the angle $\pi$ measuring the $j$-th angle exterior to $j$-th vertex of the inclusion (for instance an equilateral triangle has $\beta_1=\beta_3=\beta_5 = 5/3$ and
$\beta_0=\beta_2=\beta_4 = 1$), the following property holds true
\beq
\label{eq_somma_angoli_esterni}
\sum_{j=0}^{2n-1} \beta_j=2(n+1).
\eeq

With reference now to a isotoxal polygonal star inclusion defined by $n$ points,
the Schwarz-Christoffel conformal mapping (see \cite{tobin} and \cite{savin}) is used to
map the exterior region of the inclusion (within the physical $z-plane$) onto the exterior region of the unit circle
(within the conformal $\zeta-plane$), namely
 \beq
 \label{eq_sc_integral_ext_ext_genral}
 \barr{cll}
 \omega(\zeta)&=&\ds a \Omega \int^{\zeta}_1\left[\frac{1}{\sigma^2}\prod^{2n-1}_{j=0}\left(\sigma- k_j\right)^{\beta_{j}-1}\right]\,\mathrm{d}\sigma,
 \earr
 \eeq
where $a$ is the radius of the circle inscribing the inclusion, $\Omega$ is the scaling factor of the inclusion,
$k_j$ is the pre-image of the $j$-th polygon vertex in the $\zeta$ plane.

The first derivative of the  conformal mapping (\ref{eq_sc_integral_ext_ext_genral}) becomes
\beq
\label{eq_diff_omega_semplice}
\omega'(\zeta)=\ds a \Omega \frac{1}{\zeta^2} \prod^{2n-1}_{j=0}\left(\zeta- k_j \right)^{\beta_{j}-1}.
\eeq
Further exploiting the identity (\ref{eq_somma_angoli_esterni}), the first derivative of the conformal mapping (\ref{eq_diff_omega_semplice}) can be rewritten as
\beq
\label{eq_diff_omega}
\omega'(\zeta)=\ds a \Omega \prod^{2n-1}_{j=0}\left(1-\frac{k_j}{\zeta}\right)^{\beta_{j}-1}.
\eeq

With reference to a $n$-pointed isotoxal star polygon, see Fig. \ref{fig_star_shaped_poly_mapping}, the pre-images
and the exterior angles $\beta_{j}$ appearing in equation (\ref{eq_diff_omega}) are respectively given by
\beq
\label{preimagesptII}
k_j=e^{\i\dfrac{j\pi}{n}}
\qquad \mbox{and} \qquad
 \beta_j=
 \left\{
 \begin{array}{llll}
2(1-\xi) \qquad  &\mbox{if }  \,\, j \,\,\, \mbox{is even}\\[4mm]
2\left(\xi+1/n\right) \qquad &\mbox{if}  \,\, j \,\,\,\mbox{is odd}
 \end{array}
 \right.
 \qquad j=0,..., 2n-1,
\eeq
where $\xi$ is the fraction of $\pi$ of the semi-angle at the isotoxal-points, restricted to the following set
\beq\label{restringixi}
\xi\in\left[0,\frac{1}{2}-\frac{1}{n}\right],
\eeq
and that can be used to define the inclusion sharpness, starting from $\xi=0$, which corresponds to zero-thickness (infinite sharpness) inclusion, ending with $\xi=1/2-1/n$ corresponding to $n$-sided regular polygonal case.

\begin{figure}[tp]
  \begin{center}
\includegraphics[width=15 cm]{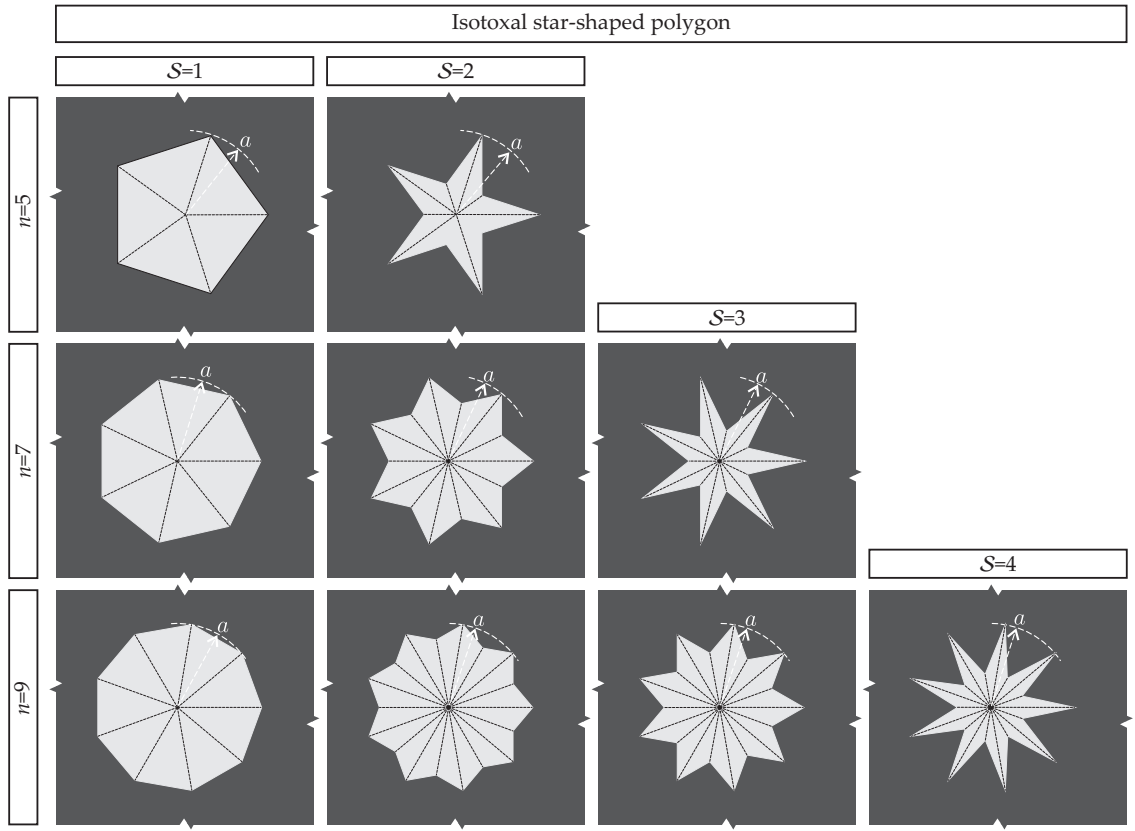}
\caption{$n$-pointed isotoxal star-shaped polygons ($n=\{5,7,9\}$, from the upper part to the lower) inscribed in a circle of radius $a$
can be used to describe inclusions in the form of $n$-sided regular polygons and $n$-pointed regular stars
with density $\mathcal{S}$ ($= \{1,2,3, 4\}$, from left to right), within an infinite elastic plane.
Note that, for a fixed $n$,
the density parameter $\mathcal{S}$ can vary only within a finite range of natural numbers $\mathcal{S}<n/2$.}
\label{settestelle}
 \end{center}
\end{figure}

\begin{figure}[tp]
  \begin{center}
\includegraphics[width=15 cm]{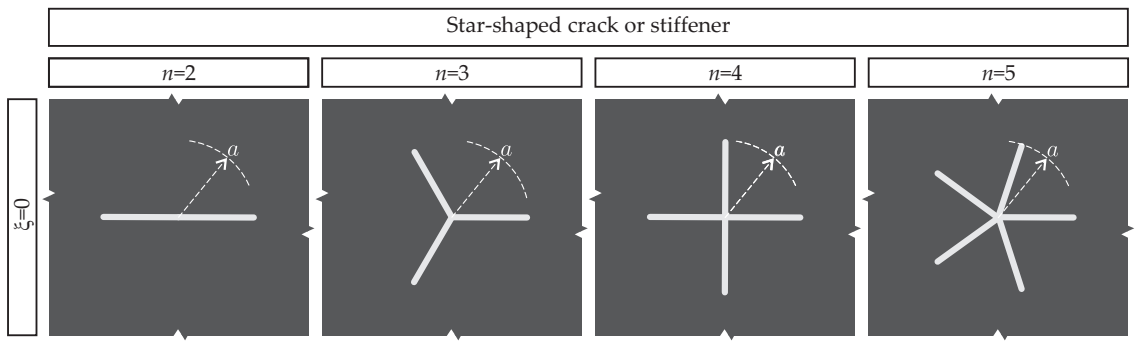}
\caption{$n$-pointed star-shaped cracks/stiffeners, obtained in the limit of $\xi=0$ of isotoxal star-shaped polygons.}
\label{ottostelle}
 \end{center}
\end{figure}

From the definition (\ref{preimagesptII})$_1$ of the pre-images $k_j$ (as the complex $n$-th roots of the positive and negative unity),
the following identities, which will become useful later, can be derived
\beq
\label{eq_roots_unity1}
\prod_{j = 0}^{n-1} \left(\zeta-k_{2j}\right)=\zeta^{n}-1, \qquad
\prod_{j = 1}^{n} \left(\zeta-k_{2j-1}\right)=\zeta^{n}+1 ,
\eeq
which can be written in an equivalent and useful form, for the future calculations, as  given below
\beq
\label{eq_roots_unity}
\prod_{j = 0}^{n-1} \left(1-\frac{k_{2j}}{\zeta}\right)=1-\frac{1}{\zeta^{n}}, \qquad
\prod_{j = 1}^{n} \left(1-\frac{k_{2j-1}}{\zeta}\right)=1+\frac{1}{\zeta^{n}} .
\eeq
The particular case of an $n$-pointed star polygon is identified  through the Schl\"{a}fli symbol $\left|n/\mathcal{S}\right|$ involving the
density, or starriness, $\mathcal{S}\in \mathbb{N}_1$, which is subject to the constraint $\mathcal{S}<n/2$, see \cite{branko, coexter}.
Therefore, for star polygons, the following relation
\beq
\xi=\frac{1}{2}-\frac{\mathcal{S}}{n} ,
\eeq
holds true, so that a regular $n$-sided polygon is recovered when $\mathcal{S}=1$, see equation (\ref{restringixi}). Now $\mathcal{S}$ controls the sharpness, so that
the higher is $\mathcal{S}$, the sharper is the star, as shown in Fig. \ref{settestelle}. In the limit case $\xi=0$ the star-shaped crack or stiffener is obtained (Fig. \ref{ottostelle}).

The generic conformal mapping (\ref{eq_sc_integral_ext_ext_genral})
can be expressed through the following Laurent series (\cite{muskio} and \cite{movchan})
\beq
\omega(\zeta) = a \Omega \left(\zeta+\sum^{\infty}_{j=1} \frac{d_{j}}{\zeta^{j}}\right) \,,
\eeq
where $d_j$ are complex constants depending on the inclusion shape.

In the following subsections,
the conformal mappings for $n$-pointed star shaped cracks and stiffeners (zero-thickness, $\xi=0$) and
isotoxal star polygonal voids or rigid inclusions (non-zero thickness, $\xi\neq0$) will be obtained as special cases of the Laurent series as
\beq
\label{conformal_map_approx}
\omega(\zeta, \xi, n) = a \Omega(n,\xi)  \sum^{\infty}_{j=0}  d_{1-j n}(\xi) \,\zeta^{1-j n} \,,
\eeq
where the scaling factor $\Omega(n,\xi)$ and the constants $d_{1-j n}(\xi)$ will be given specific expressions. It is noteworthy
that \lq $jn$'  denotes the multiplication $j\times n$ between the index $j$ and the number of points $n$.

\subsection{Star-shaped crack and stiffener}

An $n$-pointed star-shaped crack or a star-shaped stiffener can be obtained as the limit case of a isotoxal star polygon with an inifinte sharpness i.e. $\xi=0$, see Fig. \ref{ottostelle}.

Considering now a $n$-pointed regular star-shaped crack or rigid line inclusion and introducing $\xi=0$
in the definition (\ref{preimagesptII})$_2$, the first derivative of the conformal mapping (\ref{eq_diff_omega}),
together with equation (\ref{eq_roots_unity}), simplifies to
\beq
\label{eq_sc_integral_ext_ext_regular_polygon_diffrential_form_simple_stella_crack}
\omega^\prime(\zeta,n)=\ds a \Omega(n)\, \left(1-\frac{1}{\zeta^{n}}\right)\left(1+\frac{1}{\zeta^{n}}\right)^{\frac{2-n}{n}},
\eeq
where $\Omega(n)$ is a function of the number $n$ of star points given as
\beq
\label{eq_shape_parameter_poligoni_regolari_stella_crack}
\Omega(n)= \ds \frac{1}{\sqrt[n]{4}}\quad \in\,\, \left[\frac{1}{2}\,,1\right),
\eeq
for which the lower bound is obtained for $n=2$ (line inclusions, crack or stiffener) while
the upper bound corresponds to a circle, $n\rightarrow \infty$.

From the integration of  equation (\ref{eq_sc_integral_ext_ext_regular_polygon_diffrential_form_simple_stella_crack}), the mapping function
can be obtained as
\beq
\label{eq_ext_ext_map_stella_crack}
\omega(\zeta,n)=\ds  \frac{a}{\sqrt[n]{4}}\, \zeta\left(1+\frac{1}{\zeta^{n}}\right)^{\frac{2}{n}} ,
\eeq
which, using the generalized binomial theorem, can be expressed as
\beq
\label{conformal_map_approx_star_Crack}
\omega(\zeta,n) = \frac{a}{\sqrt[n]{4}} \sum^{\infty}_{j=0} \, \left[\prod_{k=0}^{j-1} \ds \left(\frac{2}{n}-k\right)\right] \,\frac{\zeta^{1-j n}}{j!} \,,
\eeq
namely, the Laurent series (\ref{conformal_map_approx}) with the complex coefficients $d_{1-j n}$
defined as
\beq
d_{1-j n} = \frac{1}{j!} \prod_{k=0}^{j-1} \ds \left(\frac{2}{n}-k\right).
\eeq

In the special case of  a simple crack or a rigid line inclusion ($n=2$),
equation (\ref{eq_ext_ext_map_stella_crack}), as well as the Laurent series (\ref{conformal_map_approx_star_Crack}),
reduces to the well-known conformal mapping function
\beq
\label{eq_SC_ext_ext_inegtral_form_line_inclusion}
\omega(\zeta)=\ds \frac{a}{2}\left(\zeta+\frac{1}{\zeta}\right).
\eeq

To derive the complex potential (\ref{pot}) for star-shaped cracks and stiffeners,
it is instrumental to introduce $t$ and $q$, functions of $m$ and $n$ as
\beq\label{tiequ}
t=\frac{2(m+1)}{n}, \qquad q=\left\lfloor \frac{m+1}{n}\right\rfloor,
\eeq
where the dependence on $m$ and $n$ is omitted for simplicity and the symbol $\lfloor \cdot \rfloor$ stands for the integer part of the relevant argument.
By means of the generalized binomial theorem, the unperturbed potential can be expressed  as
\beq
\label{eq_unperturbed_t_reali_asterisk}
g^{\infty} (\zeta,n,m)=\frac{ a^{m+1}\,\,T^{(m)}}{ 2^{t}}  \sum_{j=0}^{\infty} \left( \prod_{l=0}^{j-1} t-l\right)\, \frac{\zeta^{m+1-j n}}{j!} .
\eeq

By imposing the null traction resultant condition $F_{\stackrel{\frown}{BC}}=0$ for a crack ($\chi=1$), or the rigid-body displacement condition
$w_B=w_C$ for a rigid line inclusion ($\chi=-1$), for every pairs of points $B$ and $C$ along the boundary of the unit circle in the conformal plane,
the perturbed complex potential $g^{p(m)}$ is obtained as
\begin{dmath}
\label{eq_perturbed_potential_poly_asterisk}
g^{p} (\zeta,n,m)=\frac{a^{m+1}}{2^{t}} \left\{\chi\overline{T^{(m)}}\left[- \frac{\delta_{m+1,q n}}{q!}\prod_{l=0}^{q-1}(t-l)\,\,
+ \sum_{j=0}^{q}\,\left(\prod_{l=0}^{j-1}t-l\right)\frac{1}{j!\,\zeta^{m+1-j n}}\right]-
T^{(m)}\sum_{j=q+1}^{\infty}
\, \left(\prod_{l=0}^{j-1}  t-l\right)\frac{1}{j!\,\zeta^{j n-m-1}}\right\},
\end{dmath}
where $\delta_{m+1,q n}$ is Kronecker delta, so that \lq $qn$' is a single index corresponding to the multiplication $q \times n$ between the two indices $q$ and $n$.

The complex potential follows from the sum of the perturbed and unperturbed potentials as
\begin{dmath}
\label{eq_total_potential_poly_asterisk}
g(\zeta,n,m)=\frac{a^{m+1}}{ 2^{t}} \left\{
-\frac{\chi\overline{T^{(m)}} }{q!}\delta_{m+1,q n}\prod_{l=0}^{q-1}(t-l)+
\sum_{j=0}^{q}\,\frac{1}{j!} \left(\prod_{l=0}^{j-1} \left(t-l\right)\right)
\left[T^{(m)}  \zeta^{m+1-j n}  + \frac{\chi \overline{T^{(m)}}}{\zeta^{m+1-j n}}\right]\right\}.
\end{dmath}

Note that in the particular case when $t=2(m+1)/n \in \mathbb{N}$, the  binomial theorem can be exploited and the
complex potentials (\ref{eq_unperturbed_t_reali_asterisk}), (\ref{eq_perturbed_potential_poly_asterisk}) and (\ref{eq_total_potential_poly_asterisk}) reduce to
\beq
\label{eq_unperturbed_potential_star_crack_special_cases}
\begin{array}{lll}
\ds g^{\infty} (\zeta,n,m)=\frac{a^{m+1}\,\,T^{(m)}}{2^{t}} \sum_{j=0}^{t} \frac{t!}{j!(t-j)!} \, \zeta^{m+1-n j},\\[8mm]
\ds g^{p}(\zeta,n,m)=\frac{a^{m+1} \left(\chi \overline{T^{(m)}}- T^{(m)}\right)}{2^{t}} \left[
-\frac{t!}{q!q!} \delta_{m+1,q n}+
 \sum_{j=0}^{q} \frac{t!}{j!(t-j)!}  \frac{1}{\zeta^{m+1-j n}}\right],\\[8mm]
\ds g(\zeta,n,m)=\frac{a^{m+1}}{2^{t}} \left[
-\frac{t!}{q!q!} \chi \overline{T^{(m)}}\,\delta_{m+1,q n}
+\sum_{j=0}^{q} \frac{t!}{j!(t-j)!} \left(T^{(m)} \zeta^{m+1-n j}+
\frac{\chi \overline{T^{(m)}}}{\zeta^{m+1-n j}} \right)\right].
\end{array}\eeq

In addition to the particular case (\ref{eq_unperturbed_potential_star_crack_special_cases}),
the complex potential (\ref{eq_total_potential_poly_asterisk}) also simplifies in some other special cases, which are listed below.
\begin{itemize}
\item
$n>m+1$ (corresponding to the case $q=0$)
\beq
\label{eq_total_potential_poly_star_crack_case3_tot}
g(\zeta, n,m)=\frac{a^{m+1}}{ 2^{t}}  \left[T^{(m)} \zeta^{m+1} + \frac{\chi \overline{T^{(m)}}}{\zeta^{m+1}}\right],
\eeq
a simple expression representing an infinite set of solutions, such as that for a cruciform crack
($n=4$, Fig. \ref{ottostelle}) subject to uniform, linear and quadratic remote antiplane shear loads ($m=0,1,2$);

\item
$n=2$ (corresponding to the case of line stiffener or crack)
\beq
\label{eq_total_potential_crack}
g(\zeta,n,m)=\frac{(m+1)! \,\,a^{m+1}}{2^{m+1}} \left[
-\frac{\chi \overline{T^{(m)}}}{q!q!} \,\delta_{m+1,2q}
+\sum_{j=0}^{q} \frac{\ds T^{(m)} \zeta^{m+1-2 j}+
\frac{\chi \overline{T^{(m)}}}{\zeta^{m+1-2 j}}}{j!(m+1-j)!} \right];
\eeq
\item
$m=0$ (corresponding to the case of uniform antiplane shear \cite{sih})
\beq
g(\zeta,n)=\frac{a}{ \sqrt[n]{4}}  \left[T^{(0)} \zeta + \frac{\chi \overline{T^{(0)}}}{\zeta}\right],
\eeq
\end{itemize}
where the constant $T^{(0)}$ and its complex conjugate $\overline{T^{(0)}}$ represent the remote uniform antiplane shear loading given by the equation (\ref{eq_unperturbed_potential_1}).

\subsection{Isotoxal star-shaped polygonal voids and rigid inclusions}

Exploiting equation (\ref{eq_roots_unity1}), the first derivative of the conformal mapping (\ref{eq_diff_omega_semplice}) for an $n$-pointed isotoxal star polygon (in the case
of $\xi\neq0$) is
\beq
\label{eq_sc_integral_ext_ext_regular_polygon_diffrential_form_simple_stella_generale}
\omega^\prime(\zeta,\xi,n)=\ds a \Omega(n,\xi)
\frac{\left(\zeta^{n}-1\right)^{1-2\xi}\left(\zeta^{n}+1\right)^{2\left(\xi+\frac{1}{n}\right)-1}}{\zeta^{2}},
\eeq
where the scaling factor $\Omega$ is given by
\beq
\label{eq_shape_parameter_poligoni_regolari_stella_generale}
\Omega(n,\xi)= \frac{1}{\sqrt[n]{4}}\dfrac{\Gamma\left(1-\frac{1}{n}-\xi\right)}{\Gamma\left(\frac{n-1}{n}\right)\Gamma\left(1-\xi\right)}
\quad \in\,\, \left[\frac{1}{2}\,,1\right),
\eeq
with the symbol $\Gamma(\cdot)$ standing for Euler gamma function defined via the following convergent improper integral
\beq
\Gamma(u)=\int_{0}^{\infty} \sigma^{u-1} e^{-\sigma} d\sigma.
\eeq
Note that the lower value of $\Omega(n,\xi)$ in equation (\ref{eq_shape_parameter_poligoni_regolari_stella_generale}) is attained in the line inclusion case ($n=2$ and $\xi \rightarrow 0$),
while the upper limit is given by circle limit $n\rightarrow \infty$.

Integrating equation (\ref{eq_sc_integral_ext_ext_regular_polygon_diffrential_form_simple_stella_generale}),
it is possible to write the mapping function through Appell hypergeometric function $F_1$ \cite{abramo}, as

\beq
\label{appell}
\omega(\zeta,\xi,n)=\ds a \Omega(n,\xi) \, \zeta \,  F_1\left(-\frac{1}{n};2\xi-1,1-2\xi-\frac{2}{n};1-\frac{1}{n};\frac{1}{\zeta^n},-\frac{1}{\zeta^n}\right) ,
\eeq
which, since $\left|\zeta\right|\geq 1$, becomes
\beq
\label{appellseries}
\omega(\zeta,\xi,n)=\ds a \Omega(n,\xi) \sum_{v=0}^{\infty}\sum_{u=0}^{\infty}  \frac{\left(-\frac{1}{n}\right)_{u+v}\left(-1+2\xi\right)_{u}\left(1-2\xi-\frac{2}{n}\right)_{v}}{\left(1-\frac{1}{n}\right)_{u+v} u! v!}(-1)^{v} \zeta^{1-n(u+v)}  ,
\eeq
where, for $x\in \mathbb{R}$ and $j \in \mathbb{N}$, the symbol $(x)_{j}$ denotes the Pochhammer symbol expressed through the Euler gamma function, as
\beq
(x)_j=\frac{\Gamma(x+j)}{\Gamma(x)}.
\eeq
Transforming the index $u+v$ of equation (\ref{appellseries}) into a single index leads to
\beq
\label{appellseriessingle}
\omega(\zeta,\xi,n)=\ds a \Omega(n,\xi) \sum_{j=0}^{\infty} \sum_{k=0}^{j}
\frac{(-1)^{j-k}}{k!(j-k)!}
\frac{\Gamma\left(1-\frac{2}{n}-2\xi+j-k\right)\Gamma\left(-1+2\xi+k\right)}
{\left(1-j n\right)\Gamma\left(1-\frac{2}{n}-2\xi\right)\Gamma\left(-1+2\xi\right)}  \zeta^{1-n j} ,
\eeq
which is the Laurent series (\ref{conformal_map_approx}) with the complex constants $d_{1-j n}(\xi)$ identified as
\beq
\lb{checazzo}
d_{1-j n}(\xi)=\frac{1}{1-j n} \sum_{k=0}^{j}
\frac{(-1)^{j-k}}{k!(j-k)!}
\frac{\Gamma\left(1-\frac{2}{n}-2\xi+j-k\right)\Gamma\left(-1+2\xi+k\right)}
{\Gamma\left(1-\frac{2}{n}-2\xi\right)\Gamma\left(-1+2\xi\right)} .
\eeq
The conformal mapping (\ref{appellseriessingle}) simplifies in the following particular cases of $n$-pointed isotoxal star polygons.
\begin{itemize}
\item  $n$-sided regular polygon (so that $\xi=1/2-1/n$, with $n\geq2$), for which the scaling factor $\Omega$ and the constants $d_{1-j n}$ are
\beq
\label{eq_shape_parameter_poligoni_regolari}
\barr{lll}\Omega(n)=\ds  \frac{\sqrt{\pi} }{\sqrt[n]{4}\,\,\Gamma\left(\frac{1}{2}+\frac{1}{n}\right)\Gamma\left(1-\frac{1}{n}\right)} \quad
\in\,\left[\frac{1}{2}\,,1\right), \\[6mm]
d_{1-j n}=\ds \frac{\Gamma\left(j-\frac{2}{n}\right)}{j!(1-j n)\Gamma\left(-\frac{2}{n}\right)},
\earr
\eeq
\item  $n$-pointed regular star polygon with density $\mathcal{S}=2$ (so that $\xi=1/2-2/n$, with $n\geq 4$; for instance $n=5$ corresponds to a non-intersecting five-point star),
for which the scaling factor $\Omega$ is
\beq
\label{eq_shape_parameter_poligoni_regolari_stella}
\Omega(n)= \dfrac{\sin\left(\frac{\pi}{n}\right)}
{\pi}
\dfrac{\Gamma\left(\frac{2}{n}\right)^2 }
{\Gamma\left(\frac{4}{n}\right)} \quad \in\,\, \left[\frac{1}{\sqrt{2}}\,,1\right),
\eeq
and the coefficients $d_{1-j n}$ are
\beq
d_{1-j n}=\frac{1}{1-j n} \sum_{k=0}^{j} \frac{(-1)^{j-k}}{k!(j-k)!}
\frac{\Gamma\left(\frac{2}{n}+j-k\right)
\Gamma\left(-\frac{4}{n}+k\right)}{\Gamma\left(\frac{2}{n}\right) \Gamma\left(-\frac{4}{n} \right)}
.
\eeq
\end{itemize}

Considering the Laurent series of the mapping function (\ref{conformal_map_approx}) for the case $\xi\neq0$ with the coefficients (\ref{checazzo})
to represent the unperturbed stress (\ref{eq_general_b.c.polar}), the corresponding
unperturbed complex potential in the conformal plane can be obtained through the multinomial theorem \cite{multi} as
\beq
\label{eq_unperturbed_potential_polygono}
g^{\infty} (\zeta, \xi, n,m)= (a\Omega(n,\xi))^{m+1} \,\, T^{(m)}\sum_{j=0}^{\infty} L_{m+1-j n} \, \zeta^{m+1-j n},
\eeq
where the coefficients $L_{m+1-j n}$ are given in the form
\beq
\lb{pallissime}
L_{m+1-j n}=\sum_{\mathcal{C}_j\left(l_0, l_1,..,l_\infty\right)} \binom{m+1}{l_0, l_1,\cdots,l_\infty}  \prod_{k=0}^{\infty}  \left(d_{1-k n}\right)^{l_k} ,
\eeq
with $\mathcal{C}_j\left(l_0, l_1,..,l_\infty\right)$ representing the double conditions applied on the sum, as
\beq
\mathcal{C}_j\left(l_0, l_1,..,l_\infty\right):\left\{\sum_{k=0}^{\infty}l_k = m+1 \quad \bigcap \quad \sum_{k=1}^{\infty}k \,l_k=j \right\} ,
\eeq
where $l_k \in {\mathbb N}$.
Note that the symbol in brackets in equation (\ref{pallissime}) represents the multinomial coefficient defined through the factorial function, as
\beq
\binom{m+1}{l_0, l_1,\cdots,l_\infty}=\frac{(m+1)!}{l_0!l_1! \cdots l_\infty!}\,.
\eeq

Introduction of the boundary conditions expressing either null traction along the boundary of the star-shaped void ($\chi=1$)
or allowing only for a rigid body-displacement of the rigid inclusion along its boundary ($\chi=-1$),
the perturbed complex potential $g^{p(m)}(\zeta, \xi, n)$ is obtained as
\begin{dmath}
\label{eq_perturbed_potential_poly}
g^{p} (\zeta, \xi, n,m)=(a\Omega(n,\xi))^{m+1} \left[\chi\overline{T^{(m)}}\left(-L_{m+1-q n}\,\,\delta_{m+1,q n}+ \sum_{j=0}^{q}
\frac{L_{m+1-j n}}{\zeta^{m+1-j n}}\right)- T^{(m)}\sum_{j=q+1}^{\infty}   \frac{L_{m+1-j n}}{\zeta^{j n-m-1}}\right],
\end{dmath}
so that {\it the complex potential, solution of the isotoxal star-shaped polygonal voids and rigid inclusions, follows in a closed-form solution}
\begin{dmath}
\label{eq_total_potential_poly}
g(\zeta, \xi, n,m)=(a\Omega(n,\xi))^{m+1} \left[
-\chi\overline{T^{(m)}}L_{m+1-q n}\,\,\delta_{m+1,q n}+
\sum_{j=0}^{q} L_{m+1-j n} \left(T^{(m)} \zeta^{m+1-j n} +  \frac{\chi\overline{T^{(m)}}}{\zeta^{m+1-j n}}\right)\right] ,
\end{dmath}
as the sum of a finite number of terms.

Note that the complex potential (\ref{eq_total_potential_poly}) displays a rigid-body motion component for both the rigid inclusion and the void when  $q=(m+1)/n$.
Furthermore, the complex potential (\ref{eq_total_potential_poly}) simplifies in the following particular cases
\begin{itemize}
\item $n>m+1$ (equivalent to  $q=0$),
\beq
\label{eq_total_potential_poly_star_crack_case31_tot}
g(\zeta,\xi, n,m)=(a\Omega(n,\xi))^{m+1} \left[T^{(m)} \zeta^{m+1} + \frac{\chi \overline{T^{(m)}}}{\zeta^{m+1}}\right].
\eeq
This case embraces an infinite set of solutions, for instance a $5$-pointed isotoxal star subject to uniform, linear, quadratic and cubic remote antiplane shear load ($m=0,1,2,3$);
\item
$m=0$ (remote uniform antiplane shear)
\beq
g(\zeta, \xi, n)=a\Omega(n,\xi) \left[T^{(0)} \zeta + \frac{\chi \overline{T^{(0)}}}{\zeta}\right],
\eeq
corresponding to the solution for regular polygonal inclusions \cite{japan}).
\end{itemize}

\begin{figure}[tp]
  \begin{center}
\includegraphics[width=15 cm]{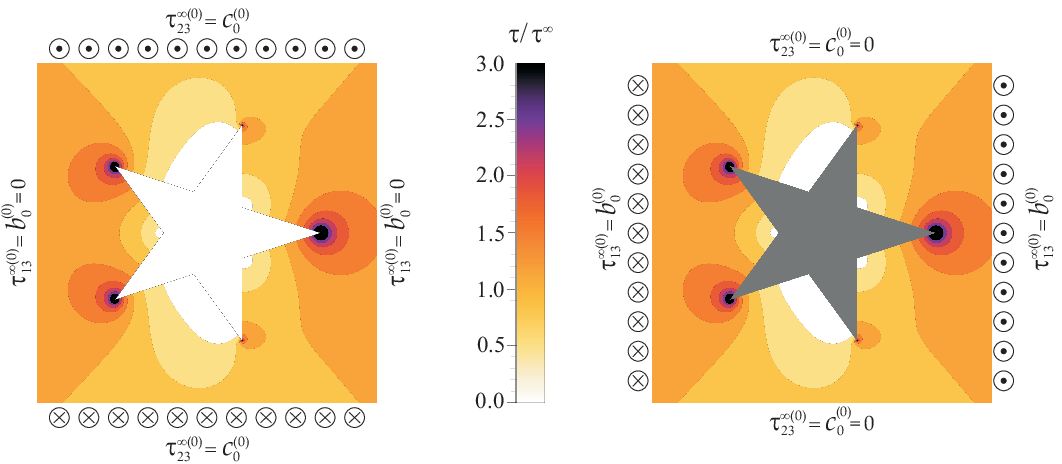}
\caption{A five pointed star ($n=5$, $\xi=1/10$) polygonal void (left) is subject to a remote uniform antiplane shear (characterized by $c_0^{\text{\tiny\ding{73}}(0)}$
and $b_0^{\text{\tiny\ding{73}}(0)}$=0), while a rigid inclusion (right) is subject to the same remote shear field but rotated of $\pi/2$
(so that the remote stress field is characterized by $b_0^{\text{\tiny\ding{72}}(0)} = c_0^{\text{\tiny\ding{73}}(0)}$ and $c_0^{\text{\tiny\ding{72}}(0)}=0$). Under these conditions, the dimensionless shear stress fields ($\tau^{(0)}(x_1, x_2)/\tau^{\infty(0)}$) are identical for both the void and the rigid inclusion.}
\label{uguali1}
 \end{center}
\end{figure}

\subsection{Shear stress analogies between rigid inclusions and voids}

The purpose of this section is to highlight some special cases in which the stress fields generated
within a matrix by a rigid inclusion are analogous to those generated when a void (of the same shape) is present.

Let us consider two remote stress fields of order $m$, equation (\ref{eq_general_b.c.polar}),
remotely applied to a matrix containing a void and a rigid inclusion
(with the same shape) and which are defined respectively by the loading constants
$ b^{\text{\tiny\ding{73}}(m)}_0, c^{\text{\tiny\ding{73}}(m)}_0$ and $b^{\text{\tiny\ding{72}}(m)}_0, c^{\text{\tiny\ding{72}}(m)}_0$.
From the obtained solution (\ref{eq_total_potential_poly}), if these constants satisfy the conditions
\beq
\lb{cazzi0}
b^{\text{\tiny\ding{73}}(m)}_0 =c^{\text{\tiny\ding{72}}(m)}_0 ,\qquad
b^{\text{\tiny\ding{72}}(m)}_0 = -c^{\text{\tiny\ding{73}}(m)}_0,
\eeq
then the following shear stress analogy occurs
\beq
\tau_{13}^{\text{\tiny\ding{73}}(m)}
\equiv
\tau_{23}^{\text{\tiny\ding{72}}(m)},
\qquad
\tau_{23}^{\text{\tiny\ding{73}}(m)}
\equiv
-\tau_{13}^{\text{\tiny\ding{72}}(m)},
\eeq
while, if the loading  constants satisfy
\beq
\lb{cazzi1}
b^{\text{\tiny\ding{73}}(m)}_0 =-c^{\text{\tiny\ding{72}}(m)}_0 ,\qquad
b^{\text{\tiny\ding{72}}(m)}_0 = c^{\text{\tiny\ding{73}}(m)}_0,
\eeq
then another shear stress analogy occurs
\beq
\tau_{13}^{\text{\tiny\ding{73}}(m)}
\equiv
-\tau_{23}^{\text{\tiny\ding{72}}(m)},
\qquad
\tau_{23}^{\text{\tiny\ding{73}}(m)}
\equiv
\tau_{13}^{\text{\tiny\ding{72}}(m)}.
\eeq
Considering the above analogies, whenever the loading constants satisfy the conditions
\beq
\lb{cazzi}
\left|b^{\text{\tiny\ding{73}}(m)}_0 \right|=\left|c^{\text{\tiny\ding{72}}(m)}_0 \right|,\qquad
\left|b^{\text{\tiny\ding{72}}(m)}_0 \right|= \left|c^{\text{\tiny\ding{73}}(m)}_0 \right|,\qquad
b^{\text{\tiny\ding{73}}(m)}_0 c^{\text{\tiny\ding{73}}(m)}_0=
-b^{\text{\tiny\ding{72}}(m)}_0 c^{\text{\tiny\ding{72}}(m)}_0,
\eeq
the modulus of the shear stress, equation (\ref{tautautau}),
within the matrix generated by the void or by the inclusion are the same
\beq
\tau^{\text{\tiny\ding{73}}(m)}
\equiv
\tau^{\text{\tiny\ding{72}}(m)}.
\eeq

An example of the identity of the fields of shear stress modulus generated by a void and a rigid inclusion under the conditions (\ref{cazzi}) is shown in Fig. \ref{uguali1} in the particular case of uniform remote stress, $m=0$.

\section{Conclusions}

Complex potentials and conformal mapping techniques have led to the analytical solution of isotoxal star-shaped polygonal
voids and rigid inclusions (and also star-shaped cracks and stiffeners) subject to remote nonuniform antiplane shear loads.
This solution will provide a guide for the development of numerical techniques in  the presence of sharp corners and is important
in the design of composite materials. The results pave the way to the discovery of situations in which the singularities
(usually present at the inclusion vertices) disappear. This important issue is systematically addressed in Part II of this study.

\section*{Acknowledgments}
The authors gratefully acknowledge financial support from the ERC Advanced Grant \lq Instabilities and nonlocal multiscale modelling of materials'
ERC-2013-ADG-340561-INSTABILITIES (2014-2019).

\end{document}